\documentclass[conference]{IEEEtran}

\usepackage{amsthm,amssymb,graphicx,multirow,amsmath,color,amsfonts}%,ulem}
\usepackage[update,prepend]{epstopdf}
\usepackage[noadjust]{cite}
\usepackage[utf8]{inputenc} % was using latin1 before
\usepackage{tikz}
\usetikzlibrary{arrows,calc}		% Optional ticks libraries
\usepackage{bbm} % for \mathbbm{1}
\usepackage{pdfpages}
\usepackage{tabulary}
\usepackage{multirow}
\usepackage{comment}
\usepackage{algorithm}
\usepackage{algpseudocode}
\usepackage{booktabs}
\usepackage{tabularx}  % For automatic column width adjustment
\usepackage{etoolbox}
\usepackage{optidef}
\usepackage[font=small,labelfont=bf]{caption}
\makeatletter

\def\nb0{{\mathbf{0}}}
\def\nb1{{\mathbf{1}}}

\allowdisplaybreaks % Allows breaking of eqnarray over multiple pages (avoids unnecessary blanks in the document before eqnarray)
\IEEEoverridecommandlockouts
\begin{document}
\graphicspath{{./Figures/}}
\IEEEaftertitletext{\vspace{-1.5\baselineskip}}
\title{
Mutual Information-Driven Visualization and Clustering for Core KPI Selection in O-RAN Testing}
%\author{\thanks{}}
\author{
    \IEEEauthorblockN{Anish Pradhan, Lingjia Liu, and Harpreet S. Dhillon}
    
    \IEEEauthorblockA{Wireless@VT, Department of Electrical and Computer Engineering, Virginia Tech, Blacksburg, VA, USA \\
    Email: \{pradhananish1, ljliu, hdhillon\}@vt.edu}
    
    \thanks{This material is based upon work supported by the National Institute of Standards and Technology (NIST), U.S. Department of Commerce, through the Public Wireless Supply Chain Innovation Fund administered by the National Telecommunications and Information Administration (NTIA), under award number 51-60-IF012. Any opinions, findings, conclusions, or recommendations expressed in this material are those of the authors and do not necessarily reflect the views of NIST, NTIA, or the U.S. Department of Commerce.}
}

 \vspace{-3mm}

\maketitle

\begin{abstract}
O-RAN testing is becoming increasingly difficult with the exponentially growing number of performance measurements as the system grows more complex, with additional units, interfaces, applications, and possible implementations and configurations. To simplify the testing procedure and improve system design for O-RAN systems, it is important to identify the dependencies among various performance measurements, which are inherently time-series and can be modeled as realizations of random processes. While information theory can be utilized as a principled foundation for mapping these dependencies, the robust estimation of such measures for random processes from real-world data remains challenging. This paper introduces AMIF-MDS, which employs aggregate mutual Information in frequency (AMIF), a practical proxy for \textit{directed information} (DI), to quantify similarity and visualize inter-series dependencies with multidimensional scaling (MDS). The proposed quantile-based AMIF estimator is applied to O-RAN time-series testing data to identify dependencies among various performance measures so that we can focus on a set of ``core'' performance measures. Applying density-based spatial clustering of applications with noise (DBSCAN) to the MDS embedding groups mutually informative metrics, organically reveals the link-adaptation indicators among other clusters, and yields a ``core'' performance measure set for future learning-driven O-RAN testing.
\end{abstract}

\section{Introduction} \label{sec:intro}
Testing and optimization in O-RAN increasingly involve handling large volumes of measurements collected from diverse subsystems, interfaces, and deployment scenarios. The open and disaggregated nature of O-RAN, while fostering innovation and multi-vendor interoperability, also leads to a rapid growth in the number of key performance indicators (KPIs) tracked during development and operation. This surge in available metrics brings both opportunity and complexity: although they hold valuable insights for diagnosing issues and refining designs, the sheer volume makes exhaustive monitoring and analysis impractical. Identifying a set of “core” KPIs, most informative for understanding and predicting network behavior, can streamline testing, simplify interoperability validation, and improve the efficiency of learning-driven O-RAN operations. Achieving this requires not only reducing the KPI set but also uncovering the structure of dependencies among the remaining metrics.

In the rapidly evolving landscape of wireless communications, these dependencies are often complex and far from intuitive. The progression towards intelligent, disaggregated architectures such as O-RAN increases the scale and intricacy of inter-KPI relationships, where both direct and indirect influences can exist. For example, the relationship between spectral efficiency (SE) and the signal-to-interference-plus-noise ratio (SINR) is nonlinear, with diminishing returns at higher SINR values. Long-term trends can also mask true dependencies, such as when interference steadily increases throughout the day as more users become active, creating temporal patterns that affect many KPIs simultaneously. These nonlinear dependencies and persistent trends can resist standard preprocessing methods, making simple correlation-based metrics insufficient \cite{spurious,spiess2010evaluation}. Untangling such relationships is essential for identifying truly central KPIs, significantly reducing the number of metrics needed for effective monitoring, streamlining complex interoperability test suites, and enabling robust feature selection in learning-driven O-RAN operations. Such insights are also vital for enhancing artificial intelligence (AI)-driven optimization in the RAN Intelligent Controller (RIC) \cite{ORAN1,ORAN2}.

Since simple metrics prove insufficient for untangling these complex dependencies and enabling the critical O-RAN operations discussed, information-theoretic measures offer a powerful framework to rigorously quantify the shared information between time-series, which are inherently realizations of random processes. Concepts such as transfer entropy  and DI are specifically designed to capture the flow of information and causal relationships between random processes. However, despite their theoretical elegance, the robust and reliable estimation of these measures analytically from continuous, real-world time-series data remains a significant challenge due to factors like finite sample sizes, noise, and the complexity of underlying dynamics \cite{bleher2024quantile}.

In light of these challenges, we propose an approach centered on AMIF, which quantifies dependencies between spectral components, thus yielding a practical and computationally feasible proxy for DI \cite{malladi2018mutual, brillinger2007mutual}. Crucially, the resulting pairwise AMIF scores can be used to construct a (dis)similarity matrix. This matrix offers a general foundation for diverse downstream tasks, including clustering, feature selection, or visualization, by capturing the information-theoretic structure among time-series. This framework is domain-agnostic and applicable to any collection of multivariate time-series. In this study, we specifically leverage this AMIF-derived similarity matrix with MDS to map and interpret these complex relationships in a reduced dimensional space. First, we validate this AMIF-MDS approach on a synthetic dataset. Then, we apply AMIF-MDS on a real-world O-RAN dataset and run a conventional clustering procedure to cluster core KPIs most suitable for feature selection in future learning-driven O-RAN operations \cite{ORAN2}. As noted before, focusing on these core metrics can also potentially identify redundant metrics and simplify interoperability test suites.

\section{Background and Approach}

To effectively analyze the intricate dependencies within time-series data, researchers have pursued various strategies. Early and classical approaches often concentrated on temporal dynamics, which is the study of how series evolve and align over time. For instance, techniques like dynamic time warping \cite{berndt1994using} assess sequence similarity by optimally aligning them in the temporal domain, accommodating phase differences. Some approaches first transform time-series into ordinal patterns to extract dynamic features before applying information-theoretic measures. Specifically, the permutation Jensen-Shannon distance \cite{zunino2022permutation} quantifies dissimilarity by comparing the distributions of ordinal patterns, effectively capturing underlying temporal structures. Similarly, the information-based correlation coefficient \cite{pernagallo2023entropy} encodes each series into dichotomous {up/down} transitions to reflect temporal dynamics, and evaluates correlation by computing the Shannon entropy of a derived matching sequence that represents the agreement in movement between the two series. As highlighted by \cite{cliff2023unifying} in their categorization of pairwise interaction statistics, it is vital to distinguish these methods, which emphasize behavioral patterns or the informativeness of specific dynamic features, from those designed to more directly quantify the overall statistical interdependency between time-series.

This distinction becomes crucial when considering the limitations of purely temporal or basic correlational approaches. Consequently, the field has increasingly turned to information-theoretic measures. For instance, the automatic mutual information optimization (AMINO) framework \cite{ravindra2020automatic} utilizes mutual information for time-series clustering. However, directly applying mutual information to raw time-series treats them as unordered sets of samples, disregarding the crucial temporal dependencies inherent in random processes. While information-theoretic measures like transfer entropy and DI are theoretically sound for capturing directed dependencies in time-series, their practical estimation from continuous data for tasks like visualization and clustering remains challenging. The authors of \cite{malladi2018mutual} introduced the concept of aggregate MIF, offering a frequency-domain perspective on information sharing. Yet, the potential of such frequency-domain information measures as comprehensive tools for visualizing and clustering time-series across diverse domains has not been fully explored, presenting a clear opportunity for advancement.

In the context of O-RAN, where thousands of performance measurements are recorded and many exhibit complex, nonlinear, and trend-driven dependencies, these same challenges and opportunities apply. Techniques capable of capturing robust information-theoretic dependencies among such time-series can directly support KPI reduction, clustering, and selection for more efficient testing and system optimization.

To address this opportunity, we developed and validated the AMIF-MDS methodology, which provides a novel and practical approach for analyzing and visualizing information-theoretic dependencies within multivariate time-series data. Our work introduces several key advancements:

\begin{itemize}
    \item We propose and implement a modified AMIF estimator. This modification replaces computationally intensive permutation testing with a significantly more efficient quantile-based approach, achieving comparable accuracy at a reduced computational cost.

    \item We integrate the AMIF estimator with multidimensional scaling by converting the AMIF-based similarity matrix into a dissimilarity matrix using membership and logarithmic transformations.

    \item We validated AMIF-MDS on synthetic time-series with known dependencies and found that it uncovers underlying relationships more accurately than traditional methods even in the presence of trends and non-linearity.

    \item We applied AMIF-MDS to visualize metrics from an O-RAN testbed dataset and then used DBSCAN on the resulting MDS embedding to identify clusters of mutually informative metrics. By interpreting these clusters, we organically revealed the link-adaptation chain.

\end{itemize}

%%%%%%%%%%%%%%%%%%%%%%
\section{Aggregate Mutual Information in Frequency}
%%%%%%%%%%%%%%%%%%%%%%
Mutual information quantifies the statistical dependence between two random variables by measuring the information-theoretic distance between two probability distributions: the actual joint distribution, which fully describes how the variables behave together, and the hypothetical distribution that represents how they would behave if they were perfectly independent. This distance is measured using the Kullback-Leibler divergence, a tool from information theory that quantifies the deviation between the true joint distribution and the product of the marginal distributions corresponding to independence. A large distance indicates a strong dependence between the variables, while a value of zero confirms that they are statistically independent.

Although mutual information provides a valuable method for detecting and quantifying dependence between random variables, numerous datasets encountered in practice do not consist of independent and identically distributed (i.i.d) samples. Such data are typically structured as dependent time-series or random processes. For data with temporal dependencies, standard mutual information proves insufficient. Extensions including mutual information rate, DI, and transfer entropy  are used to quantify information flow between random processes that account for their historical evolution. However, robust estimation of DI or transfer entropy directly from continuous data remains challenging.

To address this, MIF has emerged as a measure of statistical dependence between two random processes \cite{brillinger2007mutual,malladi2018mutual}. It quantifies the shared information between any given frequency component of the first random process and any given frequency component of the second. This is calculated via the mutual information between their respective spectral process increments derived from Cramér's spectral representation, which is conceptually related to the Fourier analysis of the processes (see e.g., \cite{malladi2018mutual} for technical details). This approach generalizes the notion of coherence for non-Gaussian processes \cite{malladi2018mutual}. Under certain conditions, aggregating MIF across relevant frequencies corresponds to the mutual information rate or even the DI between processes \cite{malladi2018mutual}, thus capturing the total information flow and directionality.

This connection highlights MIF as an interpretable, model-free tool for analyzing time-series dependence. The AMIF provides a valuable and often more tractable proxy, capturing information content between random processes. While the work in \cite{malladi2018mutual} introduced a foundational data-driven estimator for AMIF, our contribution extends significantly beyond this by developing a novel, end-to-end methodology for visualizing system-wide dependencies. To make this vision practical, we first adapted the estimation process for the specific demands of dimensionality reduction and clustering techniques. A key challenge is the computational burden of generating the comprehensive similarity matrices required for methods like multidimensional scaling. We address this by replacing the original permutation testing with a highly efficient quantile-based selection of significant frequency pairs. This is a critical first step in our broader framework, enabling the scalable application of AMIF for visualization.

\subsection{Conceptual Overview}
Building on the foundational frequency-domain analysis of mutual information \cite{brillinger2007mutual, malladi2018mutual}, the process to estimate AMIF between two time-series begins by transforming each time-series from a stream of data points into its fundamental frequency components. This is achieved by partitioning the series into smaller segments and applying a fast Fourier transform (FFT) to each one. From this frequency-domain data, a comprehensive pairwise mutual information matrix is constructed by computing the MI between every possible frequency pairing from the two series. To speed up the process, we isolate the most significant frequencies by selecting the top $q$ fraction of frequency pairs that exhibit the highest MI. The data corresponding to only these significant frequencies is then aggregated into a high-dimensional collection for each series. Finally, the mutual information is computed between these two consolidated collections and normalized, yielding a single, robust AMIF estimate that captures the essence of their statistical interdependency.

\subsection{A Walkthrough of the Estimation Process}

\begin{figure}
    \centering
    \includegraphics[clip, trim=4cm 0cm 4cm 0cm, width=\linewidth]{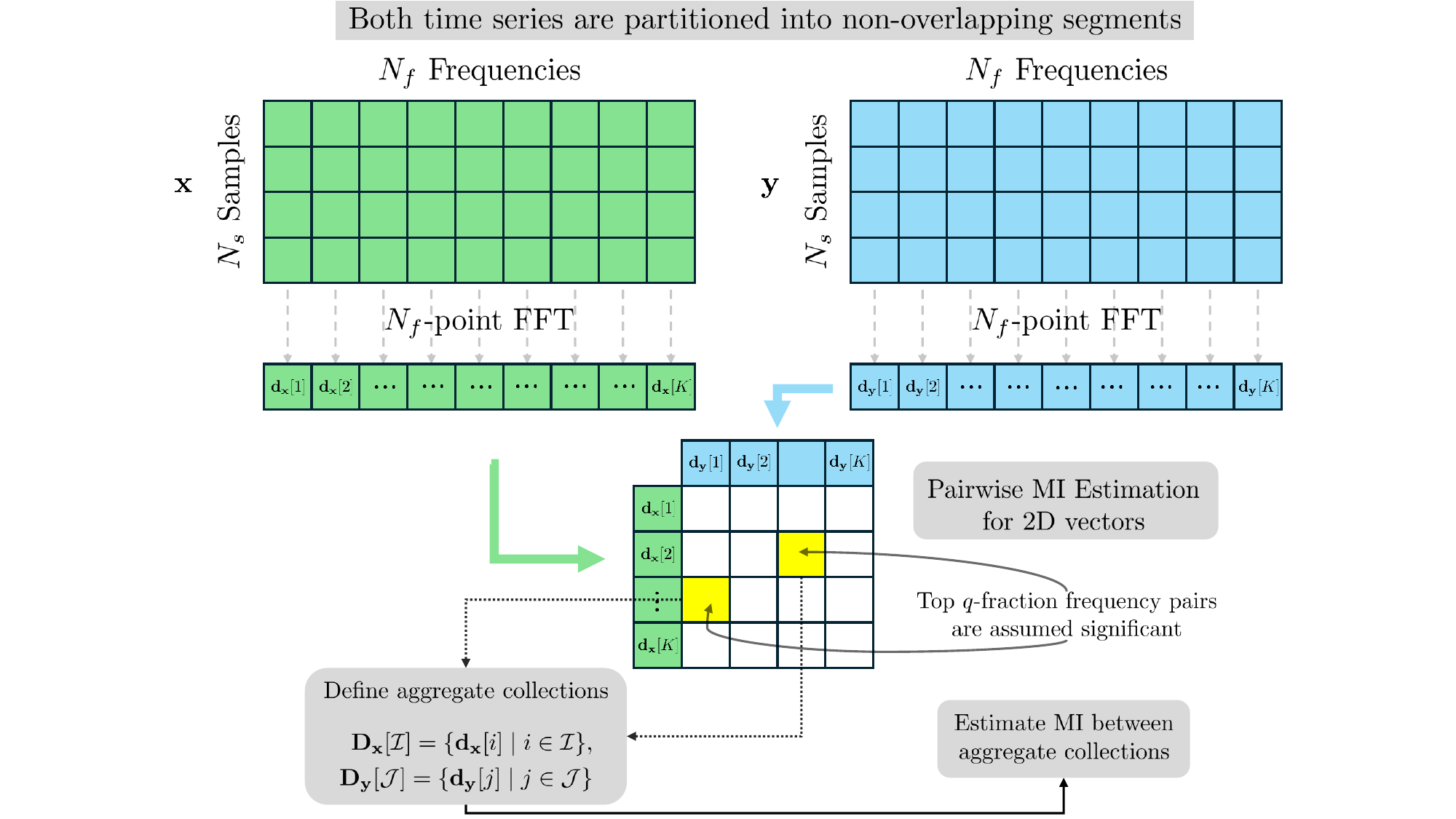}
    \caption{AMIF Estimation}
    \label{fig:amifest}
\end{figure}

Let us consider a small-scale toy example to ground these concepts, following the logic illustrated in Figure~\ref{fig:amifest}.

Suppose we have two time-series, each containing $36$ data points. The setup and deconstruction phase begins by breaking each $36$-point series into four segments of nine points each. A nine-point FFT is then applied to every segment, which converts the time-domain data into its frequency components. This initial step leaves us with four complex-valued samples for each of the nine frequencies for both series. With this frequency-domain data in hand, we then proceed to build the pairwise mutual information matrix. This involves creating a $9\times9$ matrix and filling each cell with a MI value. For instance, to get the score for the cell comparing frequency $2$ of the first series with frequency $5$ of the second series, we arrange the four complex samples for each into $4\times2$ matrices (one column for real part, one for imaginary). The MI between these two $4\times2$ matrices is then estimated using a k-Nearest Neighbors (k-NN) estimator (e.g., from the \texttt{FNN} package in R). This is repeated for all $81$ pairs to complete the matrix.

Once the matrix is built, the next stage is identifying significant frequencies and creating aggregate data. To find the most important relationships, we select the ``top-$q$ fraction" of scores. To select the top ten percent of the $81$ MI scores, we first calculate how many scores that represents. Since ten percent of $81$ is $8.1$, we round down to get a count of the top $8$ scores. This process reveals the significant frequencies involved—for instance, frequencies $\{3, 8\}$ for the first series and $\{2, 5, 9\}$ for second series. Based on these findings, we create two new aggregate collections: for the first series, we combine the $4\times2$ data matrices for its two significant frequencies to create a larger $4\times4$ matrix; for second series, we combine the data for its three significant frequencies into a $4\times6$ matrix. The final calculation involves computing a single MI score between these two new aggregate matrices, again using the k-NN estimator. After a final normalization step to ensure the score is comparable, we have our final AMIF estimate representing the shared information between the two original signals.

\section{Visualizing System-Wide Dependencies with AMIF}
The AMIF estimator detailed in the previous section provides a robust way to quantify the information shared between two time-series. To expand from this pairwise analysis to understanding the dependencies across an entire system, we can systematically apply the estimator to every pair of features in a dataset. This process allows us to construct a comprehensive similarity matrix. This matrix serves as a foundational input for various powerful visualization and clustering techniques. As our goal is to create an intuitive visualization of the system, we use this matrix with MDS to translate the complex web of dependencies into a low-dimensional map.

\subsection{Refining the Similarity Matrix for Analysis}
Once the initial pairwise AMIF scores are arranged in a matrix, two refinement steps are performed to prepare it for analysis. First, the matrix is made perfectly symmetric by averaging it with its transpose. This symmetrization is performed both to align with the formal properties of a distance metric and to satisfy the input requirements of many downstream analysis methods like clustering and visualization. Second, the diagonal elements are set to infinity, as the mutual information between a continuous random variable and itself is theoretically infinite.

%%%%%%%%%%%%%%%%%%%%%%
\subsection{Multidimensional Scaling}
To visualize the information-theoretic relationships captured by our AMIF-based similarity matrix, we employ classical MDS, a dimensionality reduction technique that preserves pairwise distances between objects in a lower-dimensional space. MDS constructs a configuration of points in the Euclidean space such that the distances between points approximate the original dissimilarities between objects. Since MDS requires a dissimilarity matrix as input rather than a similarity matrix, we define a transformation function that converts our AMIF-based similarity matrix $\mathbf{S}$ to a dissimilarity matrix $\mathbf{G}$ with the $(i,j)$-th entry denoted by $g_{ij}$ in a two-step manner. First, we normalize all the similarity scores by dividing each one by the maximum score found in the matrix. Second, we apply a function to these normalized scores to complete the conversion. Several such transformations exist, but two are particularly common. The membership transformation is straightforward: the dissimilarity is simply calculated as one minus the normalized similarity score. Another widely used method is the logarithmic transformation, where the dissimilarity is calculated as the negative logarithm of the normalized similarity. This method is especially effective at amplifying the differences between items that are very similar to each other. In our implementation, we adapt these standard functions to ensure all resulting dissimilarity scores are non-negative. For the logarithmic method, we also add a tiny constant ($\epsilon=10^{-9}$) to the similarity score before taking the logarithm to avoid any mathematical errors with values at or near zero.

\section{Synthetic Data Validation}
For rigorous testing, we create a synthetic time-series with known dependencies, nonlinearities, and trends, a design challenging traditional linear metrics that typically fail to discern the underlying structure.
\begin{algorithm}
\caption{Parent-Child AR(3) Process Generation}
\begin{algorithmic}[1]
    \Statex \textbf{Input:} The desired time series length $T$, the number of parent processes to generate $N_P$, and the trend scale factor $\alpha$.
    \Statex \textbf{Output:} A data matrix $\mathbf{D}$ of size $T \times 2N_P$ containing the generated time series, and a corresponding label vector $\boldsymbol{\ell}$.

    \State Begin by creating a time index vector $\mathbf{t}$, which is a sequence from 1 to $T$. This will be used to add a trend to the data.

    \For{each of the $N_P$ parent processes}
        \State \textbf{Generate the Parent Process:}
        \State First, sample three autoregressive coefficients ($a_1, a_2, a_3$) from a continuous uniform distribution on the interval $[-0.5, 0.5]$.
        \State Next, generate a noise vector $\boldsymbol{\varepsilon}$ of length $T$ by sampling from a standard normal distribution.
        \State Construct the parent time series $\mathbf{x}$ using an autoregressive model of order 3. For each time step $\tau$ from 4 to $T$, the value is calculated as:
        $$ x_\tau = a_1 x_{\tau-1} + a_2 x_{\tau-2} + a_3 x_{\tau-3} + \varepsilon_\tau.$$
        The initial values of the series are taken directly from the noise vector.

        \State \textbf{Add Trend and Create Child Process:}
        \State Introduce a linear trend to the parent series $\mathbf{x}$. To do this, sample a slope $\beta$ from a uniform distribution on the interval $[-\alpha, \alpha]$ and add the product of the slope and the time index vector, $\beta\mathbf{t}$, to the series $\mathbf{x}$.
        \State Create a corresponding child time series $\mathbf{y}$ by performing an element-wise square of the parent series $\mathbf{x}$.

        \State \textbf{Store and Label the Data:}
        \State Assign the newly generated parent series $\mathbf{x}$ and child series $\mathbf{y}$ to be adjacent columns in the data matrix $\mathbf{D}$.
        \State Assign the same unique integer label to both the parent and child series in the label vector $\boldsymbol{\ell}$ to mark them as a related pair.
    \EndFor

    \State After all pairs have been generated, normalize each column of the data matrix $\mathbf{D}$ to have a mean of zero and a variance of one.

    \State \textbf{return} The final data matrix $\mathbf{D}$ and the label vector $\boldsymbol{\ell}$.
\end{algorithmic}\label{alg:synth_data_gen}
\end{algorithm}

\subsubsection{Independent Parents with Nonlinear Children}
We generate a synthetic dataset comprising $N_p$ independent parent autoregressive (AR) processes, denoted as $\mathbf{x}_i$ for $i=1, \dots, N_p$. Each parent $\mathbf{x}_i$ is an AR(3) model with randomly sampled coefficients, driven by independent Gaussian noise, and incorporates a slight random linear trend component. Additionally, a corresponding child series, $\mathbf{y}_i$, is deterministically derived from each parent $\mathbf{x}_i$ through an element-wise quadratic transformation: $\mathbf{y}_i = \mathbf{x}_i \circ \mathbf{x}_i$, where $\circ$ signifies element-wise multiplication. Subsequently, all parent ($\mathbf{x}_i$) and child ($\mathbf{y}_i$) series are standardized to have zero mean and unit variance. This generation process establishes a dataset with a clear, known dependency structure:
\begin{itemize}
    \item Parent processes ($\mathbf{x}_i$) are, by construction, statistically independent of one another.
    \item Each child series ($\mathbf{y}_i$) is a deterministic nonlinear function solely of its respective parent ($\mathbf{x}_i$).
    \item Consequently, parent-child pairs from different generative families ($\{\mathbf{x}_i, \mathbf{y}_i\}$ vs. $\{\mathbf{x}_j, \mathbf{y}_j\}$ for $i \neq j$) are expected to form distinct clusters.
\end{itemize}

This synthetic dataset deliberately introduces two significant challenges for traditional time-series analysis: a) random linear trends that can induce spurious correlations between otherwise independent series \cite{spurious}, and b) nonlinear relationships through squared transformations. While conventional correlation-based approaches may fail to accurately identify true dependencies in such data-either being misled by trend-induced correlations or missing nonlinear relationships even after detrending-our method aims to overcome both obstacles simultaneously. By demonstrating the effectiveness of our approach on this controlled dataset, we can verify its ability to capture genuine information sharing between variables regardless of trends and nonlinearities, without requiring preprocessing steps like detrending that might be difficult to apply optimally in real-world scenarios.

\subsubsection{Expected Outcomes}

Using this synthetic data generation approach, we expect the time-series to form distinct clusters corresponding to parent-child pairs. These relationships should not be fully discernible using only Euclidean distance or linear correlation metrics due to the nonlinear transformation. Our proposed approach should reveal these clusters by capturing the complex statistical dependencies that traditional metrics miss. The complete algorithm for generating this synthetic dataset is presented in Algorithm \ref{alg:synth_data_gen}.

\subsection{Synthetic Data Results}
We generate sixteen series by squaring eight independent AR(3) parent processes, add random trends with $\alpha=10^{-3}$, and normalize each to zero mean and unit variance. The goal is to recover the eight parent-child pairs as distinct clusters.

\begin{figure*}
    \centering
    \includegraphics[clip, trim=0.3cm 11.4cm 0.6cm 0.2cm, width=1\linewidth]{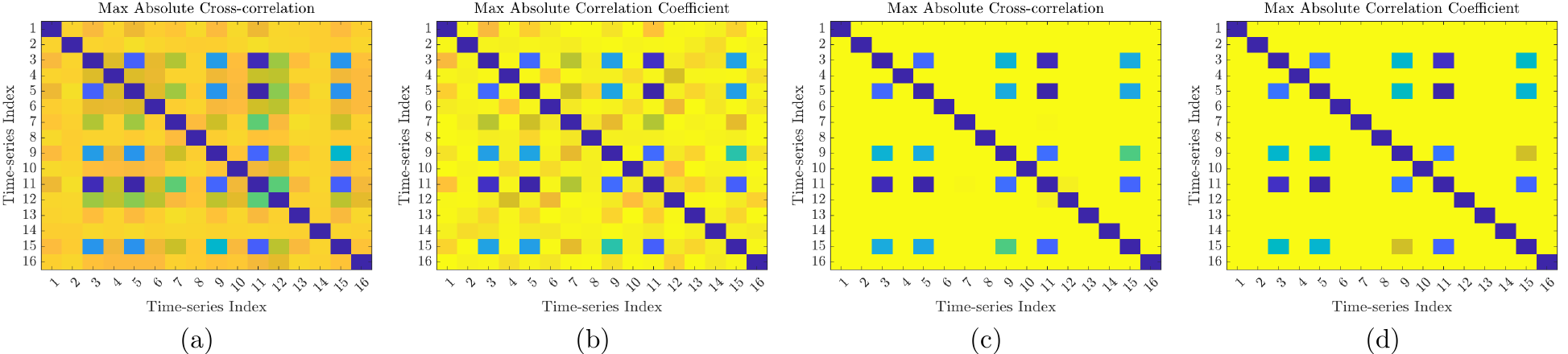}
    \caption{Dissimilarity matrices with membership transformation: (a) max absolute cross-correlation, (b) maximum absolute correlation coefficient, and with logarithmic transformation: (c) max absolute cross-correlation, (d) maximum absolute correlation coefficient.}
    \label{fig:Othersynthdissim}
\end{figure*}

In the Fig. \ref{fig:Othersynthdissim}, the dissimilarity matrices derived from conventional linear metrics, namely maximum absolute cross-correlation (MACC) \cite{abscrosscorr} and maximum absolute correlation coefficient (MACCoeff) \cite{abspearson}, are presented. These results, shown for both membership and logarithmic transformations, illustrate the failure of these metrics to reveal the underlying block-diagonal structure corresponding to the parent-child pairs. This is attributed to the nonlinear dependencies and induced trends inherent in the synthetic data.

\begin{figure*}
    \centering
    \includegraphics[clip, trim=0.3cm 11.4cm 0.5cm 0.2cm, width=1\linewidth]{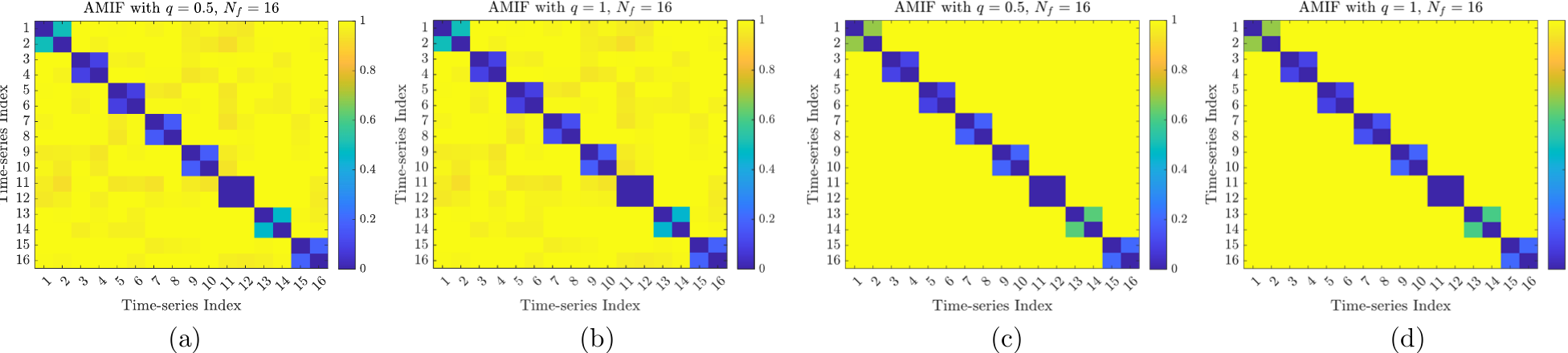}
    \caption{AMIF-based Dissimilarity matrices with membership transformation: (a) $q=0.5,\,N_f=16$, (b) $q=0.5,\,N_f=32$, (c) $q=1,\,N_f=16$, and (d) $q=1,\,N_f=32$ while the figures from (e) to (h) correspond to the logarithmic transformation.}
    \label{fig:AMIFSynthDissim}
\end{figure*}

In contrast, Fig. \ref{fig:AMIFSynthDissim} displays the dissimilarity matrices obtained using the proposed AMIF measure with various parameter settings ($q \in \{0.5, 1\}$, $N_f = 16$) and both membership (Fig. \ref{fig:AMIFSynthDissim}(a)-(b)) and logarithmic (Fig. \ref{fig:AMIFSynthDissim}(c)-(d)) transformations. The AMIF-based dissimilarity matrices consistently exhibit a pronounced block-diagonal structure, accurately identifying all eight parent-child clusters irrespective of the tested AMIF parameters or transformation type. Note that, the logarithmic transformation yielded dissimilarity values larger than unity for unrelated pairs.

\begin{figure*}
    \centering
    \includegraphics[clip, trim=0.3cm 11cm 0.4cm 0.2cm, width=1\linewidth]{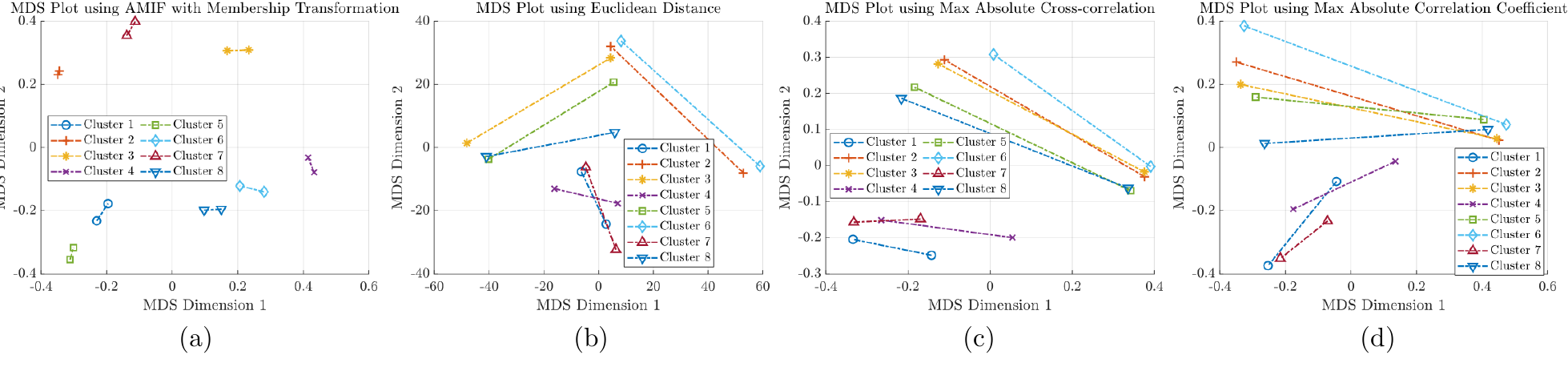}
    \caption{MDS Visualization with memebership transformation with different similarity measures: (a) AMIF, (b) Euclidean distance, (c) maximum absolute cross-correlation, and (d) maximum absolute correlation coefficient.}
    \label{fig:MDSAMIFAll}
\end{figure*}
We assess these dissimilarities via classical MDS in Fig. \ref{fig:MDSAMIFAll}, using membership‐transformed similarities for direct interpretability. In Fig. \ref{fig:MDSAMIFAll}(a), AMIF‐based dissimilarities yield distinct, well‐separated parent-child clusters, whereas the Euclidean (b), MACC (c) and MACCoeff (d) versions fail to resolve the true clusters. This confirms the superior ability of AMIF to uncover complex, nonlinear relationships in multivariate time-series for visualization.

\section{O-RAN Dataset Visualization and Discussion} 

\begin{figure}
    \centering
    \includegraphics[width=0.8\columnwidth]{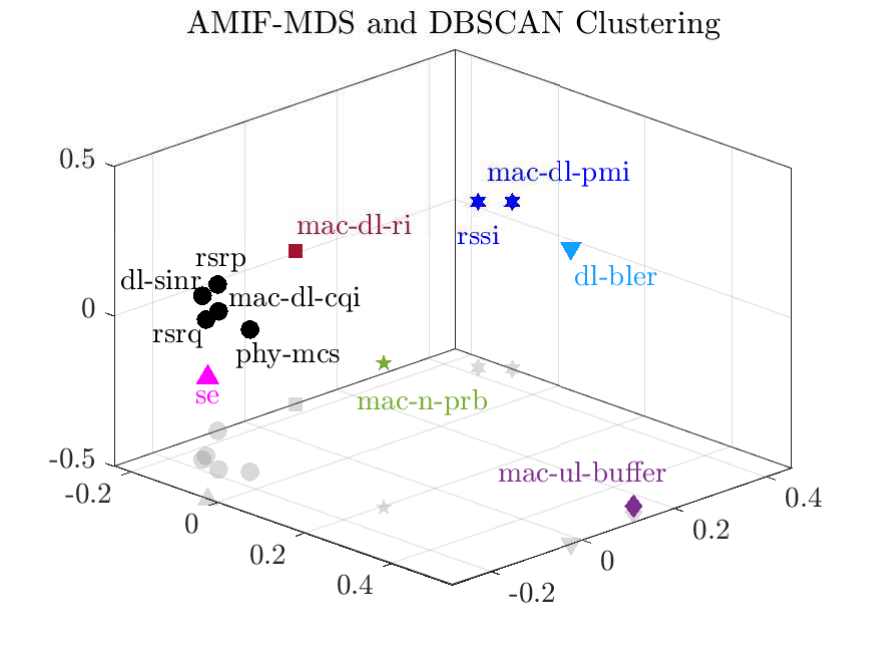}
    \caption{AMIF-MDS with membership transformation and DBSCAN.}
    \label{fig:ORANMDSDBSCAN}
\end{figure}

In this section, we analyze time series data from an O-RAN testbed subjected to random OFDM burst interference~\cite{Dai2024Learning}, and show that our method extracts the core KPIs by disentangling their dependencies. The dataset comprises thirteen PHY/MAC-layer KPIs, sampled every 20 ms (the `Packet Delay' KPI was excluded due to data incompleteness. Further details on each KPI are available in~\cite{Dai2024Learning}). To analyze inter-metric dependencies under dynamic interference, we computed an AMIF-based dissimilarity matrix with membership transformation, projected it via MDS, and applied DBSCAN  (with a neighborhood radius of $0.15$ and a minimum cluster size of $1$) to cluster metrics by shared information. Fig. \ref{fig:ORANMDSDBSCAN} presents these clusters, interpreted using domain knowledge. Note that the 3D plot employs shadows for each marker on the bottom plane, enhancing depth perception and aiding the reader in discerning the spatial relationships within the 2D representation.

The largest cluster comprises key downlink link-adaptation metrics: channel quality indicator (MAC-DL-CQI), downlink SINR (DL-SINR), reference signal received power/quality (RSRP/RSRQ), and modulation and coding scheme (PHY-MCS). These metrics form a tightly coupled chain: physical layer measurements (RSRP, RSRQ, DL-SINR) inform the user-reported CQI, which, in turn, directly governs PHY-MCS selection. Such strong causal relationships and resultant high mutual information explain their co-clustering. In contrast, SE forms a singleton cluster. This highlights the fact that SE is shaped by both these link-adaptation metric PHY-MCS and scheduling parameter MAC-N-PRB. This dual dependency is visually corroborated by the proximity of SE to these distinct metrics in the 3D plot.

DBSCAN also groups the received signal strength indicator (RSSI) with the precoding matrix indicator (MAC-DL-PMI). We believe that in our testbed, high-power burst jamming produces rapid RSSI spikes that directly reflect interference events. At the same time, MAC-DL-PMI captures the spatial processing adjustments of the user under these distorted channel conditions. The co-clustering suggests that interference peaks and PMI updates may be closely synchronized.

The remaining metrics each form singleton clusters, reflecting largely orthogonal information. The rank indicator (MAC-DL-RI) reports the channel rank. This value is independent of scalar power or SINR measurements. MAC-UL-Buffer measures uplink queue occupancy and shares minimal information with downlink physical-layer metrics. MAC-N-PRB indicates the number of allocated PRBs. It lies near the link-adaptation cluster because PRB allocation influences throughput. However, it is labeled separately as it reflects eNB scheduler decisions rather than direct channel conditions. Finally, the block error rate (DL-BLER) quantifies packet errors. Robust error correction coding largely decouples its behavior from SINR fluctuations or scheduler settings.

\section{Conclusions}
This paper introduced AMIF-MDS for visualizing and clustering O-RAN performance measurements based on their information-theoretic dependencies, although the method can be applied much more generally to any collection of time-series. The methodology integrates the proposed fast quantile-based AMIF estimator with MDS via transformed similarity matrices. On synthetic data with known nonlinearities and trends, AMIF-MDS demonstrated superior recovery of the known dependency structure compared to traditional linear metrics, providing confidence in its robustness. We then applied the method to a real-world dataset from a virtualized, disaggregated O-RAN platform, chosen because its architecture logs a multitude of tightly interrelated KPIs and therefore provides an ideal testbed for our approach. Applying DBSCAN to the MDS embedding organically revealed the link-adaptation chain and other non-trivial relationships, such as the co-clustering of RSSI and PMI, yielding a core KPI set for future learning-driven O-RAN operations and reducing the number of interoperability tests for O-RAN. These results open promising avenues for future research, including exploring alternative clustering and visualization techniques for AMIF-derived matrices, leveraging AMIF for feature selection and dimensionality reduction, and developing novel distance transformations.
%%%%%%%%%%%%%%%%%%%%%%

\bibliographystyle{IEEEtran}
\bibliography{hokie}
\end{document}